\documentclass[aps,eqsecnum]{revtex4}
\usepackage{graphicx,epsfig}
\begin{document}
\def\ii{\'{\i}}
\title{Kaon condensation in the quark-meson coupling model and compact
stars}

\author{D.P. Menezes}
\affiliation{Depto de F\ii sica - CFM - Universidade Federal de Santa
Catarina  Florian\'opolis - SC - CP. 476 - CEP 88.040 - 900 - Brazil}
\affiliation{School of Physics - University of Sydney - NSW - 2006 - Australia}
\author{P.K. Panda}
\affiliation{Depto de F\ii sica - CFM - Universidade Federal de Santa
Catarina  Florian\'opolis - SC - CP. 476 - CEP 88.040 - 900 - Brazil}
\affiliation{Centro de F\ii sica Te\'orica - Dep. de F\ii sica -
Universidade de Coimbra - P-3004 - 516 Coimbra - Portugal}
\author{C. Provid\^encia}
\affiliation{Centro de F\ii sica Te\'orica - Dep. de F\ii sica -
Universidade de Coimbra - P-3004 - 516 Coimbra - Portugal}

\begin{abstract}
The properties of neutron stars constituted of a crust of hadrons and
an internal part of hadrons and kaon condensate are calculated within the
quark-meson-coupling model. 
We have considered stars with nucleons
only in the hadron phase and also stars with hyperons as well. The results
are compared with the ones obtained from the non-linear Walecka model for the
hadronic phase.
\end{abstract}
\date{\today}
\maketitle
\vspace{0.5cm}
PACS number(s): 26.60.+c,24.10.Jv, 21.65.+f,95.30.Tg
\vspace{0.5cm}

\section{Introduction}

Neutron stars are stellar objects produced as a result of supernova explosions.
They are highly condensed since their  masses are of the order of one to two 
times the solar mass, but their radii are only 10 to 12 Km. When a neutron
star is born its temperature is of the order of $10^{11}$ K but in a few days
it cools down to about $10^{10}$ K by emitting neutrinos. The temperature at
the surface is about one-tenth that of the interior temperature and it was
found that it radiates x-rays. Neutron stars are normally treated as zero
temperature stellar objects because their temperatures, although high, are 
still very low as compared with their characteristic energies of excitation. 
Unfortunately the x-ray emissions are still not well understood to give a 
clear picture of the possible interior of the neutron stars. Their outer crust
is believed to be formed mainly by nuclei and electrons and, as density 
increases free neutrons are also expected to appear. The inner core of the
neutron stars however are still a source of expeculation, some of the 
possibilities being the appearance of hyperons, of a mixed phase of hyperons and
quarks, only quarks or pion and kaon condensates.

The equation of state (EOS) and the internal constitution of the
neutron stars depend on the nature of the strong interaction. One
possibility is the existence of matter with strangeness to baryon
ratio, $S/B$, of order unity. In a compact stars strangeness may
occur in the form of fermions such as $\Lambda$ and $\Sigma^-$
hyperons, as a Bose condensate i.e. $K^-$ meson condensate or in the
form of strange quarks in the mixed phase of hadrons and quarks or
in a pure quark phase. In the hadron phase, hyperons influence the
neutron star structure \cite{prak97,Glen00}. Keplon and Nelson
\cite{kaplan} have suggested that, above some critical density, the
ground state of baryonic matter might contain a Bose-Einstein
condensate of negatively charged kaons. There is a strong attraction
between $K^-$ mesons and baryons which increases with density and
lowers the energy of the zero momentum state. A condensate is formed
when this energy equals the kaon chemical potential $\mu$. 
Hence, in neutron stars charge neutrality favors kaon condensation 
because the neutrons at the top of the Fermi sea decay into protons plus 
electrons, which in turn, have an increase in the energy as the density 
increases. When the electron chemical potential equals the effective kaon mass,
the kaons are favored in helping with the conservation of charge neutrality 
because they are bosons and can condense in the lowest energy state. For this
reason $K^-$ mesons are the ones normally included in the calculations since
their charge is the same as the electric charge.

Many calculations for EOSs in dense matter show that the hyperons
appear at densities above $\sim~ (2-3)\rho_0$ where $\rho_0$ denote the
equilibrium nuclear matter density. Typically, the critical density for
condensation is $\sim~ (3-4)\rho_0$ in nucleon matter (although it is
model and parameter dependent). A density of this order is smaller than the
central density of the neutron star, so a $K^-$ condensate could
be present in its core.
The possibility of kaon condensation in the dense
interior of neutron stars was investigated with the help of a chiral
\cite{kaplan} and   the non-linear Walecka model (NLWM) \cite{walecka} for the
description of the hadronic phase \cite{thorsson,ellis,gl99}. In
\cite{ellis}, the baryonic octet was also included. In \cite{pons}
results obtained within the Zimani and Moskowski (ZM) model
\cite{ZM} at zero and finite temperature, have also been discussed.

In previous works, we have extensively discussed many possible EOS used to
describe hybrid stars either after the deleptonization era \cite{mp,pmp}
or before it has taken place, when neutrinos are still trapped inside the
star \cite{trapping}. The hybrid stars are constituted of a hadronic phase,
present in the outer and crust parts of the star, a mixed phase of hadrons and
quarks and possibly a central part composed of quarks only. In the afore
mentioned works we have always followed a Gibbs prescription in the
construction of the mixed phase.
The implementation of Gibbs criteria for phase coexistence requires a global
charge neutrality condition and results in a smooth EOS where no
discontinuities are present. Both
independent neutron and electron chemical potentials are enforced to be
identical in both hadron and quark phases at the phase transition
point. 

In the present work, we investigate the phase transition from a
hadronic phase to a phase with baryons and kaon condensate using the
quark-meson-coupling model (QMC) \cite{guichon,st} including
hyperons in order to describe the hadron phase. In the QMC model,
baryons are described as a system of non-overlapping MIT bags which
interact through the effective scalar and vector mean fields, very
much in the same way as in the Walecka model. Nevertheless, the
quark degrees of freedom are explicitly taken into account and the 
coupling constants are defined at the quark level, what
makes the QMC a richer model mainly when the properties of mesons in
nuclear medium are investigated. Here we treat the kaons as MIT bags 
and the couplings of kaons to mesons are determined from the formalism.

In \cite{tsushima98} the in-medium kaon and anti-kaon
properties were studied within the QMC framework. Next we use their
prescription to find the kaon properties namely, the bag
radius in medium, the effective mass of the baryons, the coupling of kaons with
hyperons.

\section{The quark-meson coupling model for hadronic matter with kaons}

In the QMC model, the nucleon in nuclear medium is assumed to be a
static spherical MIT bag in which quarks interact with the scalar and
vector fields, $\sigma$, $\omega$ and $\rho$ and these
fields are treated as classical fields in the mean field
approximation.  The quark field, $\psi_q(x)$, inside the bag then
satisfies the equation of motion:
\begin{equation}
\left[i\,\rlap{/}\partial-(m_q^0-g_\sigma^q\, \sigma_0)
-\gamma^0~(g_\omega^q\, \omega_0\, + \frac{1}{2} g^q_\rho \tau_{3q}
b_{03})\right] \,\psi_q(x)=0\ , \quad  q=u,d,s, \label{eq-motion}
\end{equation}
where $\sigma_0$, $\omega_0$ and $b_{03}$ are the classical meson
fields for sigma, omega and rho mesons. $m_q^0$ is the current quark
mass, $\tau_{3q}$ is the third component of the Pauli matrices and
$g_\sigma^q$, $g_\omega^q$ and $g_\rho^q$ are the quark couplings
with $\sigma$, $\omega$ and $\rho$ mesons. The normalized ground
state ($s$- state) for a quark in the bag is given by
\cite{guichon,st}
\begin{equation}
\psi_q({\bf r}, t) = {\cal N}_q \exp \left(-i\epsilon_q t/R_B \right)
\left( \matrix{ j_0\left(x_q r/R_B\right)\cr
i\beta_q \vec{\sigma} \cdot \hat r j_1\left(x_q r/R_B\right) }
\right) \frac{\chi_q}{\sqrt{4\pi}} ~,
\end{equation}
where
\begin{equation}
\epsilon_q=\Omega_q +R_B\left(g_\omega^q\, \omega+ \frac{1}{2}
g^q_\rho \tau_{3q} b_{03} \right)  ~; ~~~
\beta_q=\sqrt{{\Omega_q-R_B\, m_q^*\over \Omega_q\, +R_B\, m_q^*}}\
,
\end{equation}
with $\Omega_q\equiv \sqrt{x_q^2+(R_B\, m_q^*)^2}$,
$m_q^*=m_q^0-g_\sigma^q\, \sigma$, $R_B$ is the bag radius of the
baryon, $B$, $\chi_q$ is the quark spinor and ${\cal N}_q$ is the
normalization constant. The quantities $\psi_q,\, \epsilon_q,\,
\beta_q,\, {\cal N}_q,\, \Omega_q,\, m^*_q$ all depend on the baryon
considered. The boundary condition at the bag surface is given by
\begin{equation}
i\vec \gamma \cdot \hat n \psi_q=\psi_q  ~, \label{bun-con}
\end{equation}
which reduces to $j_0(x_q)=\beta_q\, j_1(x_q)$ for the ground state
and determines the dimensionless quark momentum $x_q$. The energy of
a static bag describing baryon $B$, consisting of three ground state
quarks, can be expressed as
\begin{equation}
E^{\rm bag}_B=\sum_q n_q \, {\Omega_q\over R_B}-{Z_B\over R_B}
+{4\over 3}\,  \pi \, R_B^3\,  B_B\ ,
\label{ebag}
\end{equation}
where $Z_B$ is a parameter which accounts for zero-point motion
and $B_B$ is the bag constant. The set of parameters used in the present work
is given in \cite{pmp} for the bag value $B_B^{1/4}=210.854$ MeV,
$m_u^0=m_d^0=5.5$ MeV and $m_s^0=150$ MeV.
The effective mass of a nucleon bag at rest
is taken to be
\begin{equation}
M_B^*=E_B^{\rm bag}.
\label{eff-mn}
\end{equation}
As in \cite{tsushima98}, we assume that the kaons are described by
the static MIT bag, in the same way as the nucleon and the hyperons. Moreover, the
$\sigma$, $\omega$ and $\rho$ mesons are only mediators of the $u$
and $d$ quarks inside the kaons. Similarly the effective mass for
the kaon is given by
\begin{equation}
m^*_K=\frac{\Omega_q+\Omega_s}{R_K} -\frac{Z_K}{R_K}+ \frac{4}{3}\pi
{R_K}^3 B_K, \hspace{0.5in} (q\equiv u,d)\label{mks}
\end{equation}
For our calculation, we have fixed the bag constant, $B_K$, same as
for the nucleon and from the kaon mass and the stability condition
at vacuum, we have $Z_K=3.362$ and $R_K=0.457$ fm for $R_N=0.6$ fm. 
The equilibrium condition for the bag
is obtained by minimizing the effective masses $M_B^*$ and $m^*_K$
with respect to the bag radius
\begin{equation}
\frac{\partial M_B^*}{\partial R_B^*} = 0
\label{balance}
\end{equation}
and
\begin{equation}
\frac{\partial m_K^*}{\partial R^*_K} = 0. \label{balancek}
\end{equation}

For the QMC model with kaons, the equations of motion for the meson
fields in uniform static matter are given by
\begin{equation}
m_\sigma^2\sigma = \sum_B g_{\sigma B} C_B(\sigma) \frac{2J_B +
1}{2\pi^2} \int_0^{k_B} \frac{M_B^*(\sigma)} {\left[k^2 + M_B^{*
2}(\sigma)\right]^{1/2}} \: k^2 \ dk + g_{\sigma K}~ \rho_K ,
\label{field1}
\end{equation}
\begin{equation}
m_\omega^2\omega_0 = \sum_B g_{\omega B} \left(2J_B + 1\right) k_B^3
\big/ (6\pi^2) - g_{\omega K}~ \rho_{K} , \label{field2}
\end{equation}
\begin{equation}
m_\rho^2 b_{03} = \sum_B g_{\rho B} I_{3B} \left(2J_B + 1\right)
k_B^3 \big/ (6\pi^2) - g_{\rho K} ~\rho_{K} . \label{field3}
\end{equation}
In the above equations $J_B$, $I_{3B}$ and $k_B$ are respectively
the spin, isospin projection and the Fermi momentum of the baryon
species $B$ and $\rho_{K}$ is the kaon density. In Eq.
(\ref{field1}) we have
\begin{equation}
g_{\sigma B}C_B(\sigma) = - \frac{\partial M_B^*(\sigma)}{\partial
\sigma} = - \frac{\partial E^{\rm bag}_B}{\partial \sigma} =
\sum_{q=u,d} n_q g^q_\sigma S_B(\sigma)
\end{equation}
where
\begin{equation}
S_B(\sigma) = \int_{bag} d{\bf r} \ {\overline \psi}_q \psi_q =
\frac{\Omega_q/2 + R_Bm^*_q(\Omega_q - 1)} {\Omega_q(\Omega_q -1) +
R_Bm_q^*/2} ~; ~~~~ q \equiv (u,d) ~.
\end{equation}

After self-consistent calculation, the kaon effective mass, $m^*_K$
can be parametrized as \cite{tsushima98}
\begin{equation}
m^*_K=m_K-g_{\sigma K}(\sigma) \sigma \simeq m_K-\frac{1}{3} g_{\sigma N}
\left(1-\frac{a_K}{2}g_{\sigma N} \sigma \right) \sigma,
\end{equation}
where $a_k=0.00045043\mbox{ MeV}^{-1}$ for $R_N=0.6$ fm.

Boosting the $K^\pm$  bags we get the dispersion relation for the kaons
\begin{equation}
\left(\begin{array}{c}\omega_{K^+} \\ \omega_{K^-}\end{array}
\right)=\sqrt{{m_K^*}^2+\mathbf{k}^2} \pm (g_{\omega K}\omega_0 +
\frac{1}{2}g_{\rho K}\tau_{3q} b_{03}) \label{k+-}.
\end{equation}

The above prescription is equivalent to using, for the kaon sector,
the effective lagrangian density \cite{ellis,gl99}
\begin{equation}
{\cal L}_K={\cal D}_{\mu}^*K^* {\cal D}^{\mu}K - m^*_K K^*K,
\end{equation}
with
\begin{equation}
{\cal D}_{\mu}=\partial_{\mu}+i g_{\omega K} \omega_{\mu}+
i\frac{1}{2} g_{\rho K} ~ \vec \tau\cdot {\mathbf b}_{\mu}.
\end{equation}
and the effective kaon mass defined  by (\ref{mks}).

We have used \cite{pmp} $g_\sigma^q=5.957$, $g_{\sigma
N}=3g_\sigma^q S_N(0)=8.58$, $g_{\omega N}=8.981$, $g_{\rho
N}=8.651$ with $g_{\omega N} = 3g_\omega^q$ and $g_{\rho N} =
g_\rho^q$. We have taken the standard values for the meson masses,
$m_\sigma=550$ MeV, $m_\omega=783$ MeV and $m_\rho=770$ MeV. The
hyperon couplings are not relevant to the ground state properties of
nuclear matter, but information about them can be available from the
levels in $\Lambda$ hypernuclei \cite{chrien}. In this work we have used
$g_{\sigma B}=x_{\sigma B}~ g_{\sigma N},~~g_{\omega B}=x_{\omega B}~
g_{\omega N}, ~~g_{\rho B}=x_{\rho B}~ g_{\rho N}$ and $x_{\sigma
B}$, $x_{\omega B}$ and $x_{\rho B}$ are equal to $1$ for the
nucleons and equal to $\sqrt{2/3}$ for the other baryons
\cite{moszk}. Note that the $s$-quark is unaffected by the sigma,
omega  and rho mesons i.e. $g_\sigma^s=g_\omega^s=g_\rho^s=0\ .$ The kaon
couplings are given by $g_{\omega K}=\frac{1}{3}g_{\omega N}$, $g_{\rho
K}=g_{\rho N}$

The total energy density and the pressure including the leptons and
the kaons can be obtained from the grand canonical potential. The
kaons do not contribute directly to the pressure as they are in a
$s$-wave Boson condensate, but do contribute to the energy density.
The leptons are included as a free Fermi gas since they only
interact with the baryons via the weak interaction. We then obtain:

\begin{eqnarray}
\varepsilon &=& \frac{1}{2}m_\sigma^2 \sigma^2
+ \frac{1}{2}m_\omega^2 \omega^2_0
+ \frac{1}{2} m_\rho^2 \rho^2_{03} \nonumber\\
&+& \sum_B \frac{2J_B +1}{2\pi^2} \int_0^{k_B}k^2 dk
\left[k^2 + M_B^{* 2}(\sigma)\right]^{1/2}
+ \sum_l \frac{1}{\pi^2} \int_0^{k_l} k^2  dk\left[k^2 + m_l^2\right]^{1/2}
+ \varepsilon_K,
\end{eqnarray}
where
\begin{equation}
\varepsilon_K=m^*_K \rho_K,
\end{equation}
\begin{eqnarray}
P &=& - \frac{1}{2}m_\sigma^2 \sigma^2
+ \frac{1}{2}m_\omega^2 \omega^2_0
+ \frac{1}{2} m_\rho^2 \rho^2_{03} \nonumber\\
&+& \frac{1}{3} \sum_B \frac{2J_B +1}{2\pi^2} \int_0^{k_B}
\frac{k^4 \ dk}{\left[k^2 + M_B^{* 2}(\sigma)\right]^{1/2}}
+ \frac{1}{3} \sum_l \frac{1}{\pi^2} \int_0^{k_l} \frac{k^4  dk}
{\left[k^2 + m_l^2\right]^{1/2}} ~.
\end{eqnarray}

The lepton Fermi momenta are the positive real solutions of
$(k_e^2 + m_e^2)^{1/2} =  \mu_e$ and
$(k_\mu^2 + m_\mu^2)^{1/2} = \mu_\mu = \mu_e$. The equilibrium composition
of the star is obtained by solving the set of Eqs. (\ref{field1})-
(\ref{field3}) in conjunction
with the charge neutrality condition  (\ref{neutral})  at a given total
baryonic density
$\rho = \sum_B (2J_B + 1) k_B^3/(6\pi^2)$; the baryon and kaon effective
masses are obtained self-consistently in the bag model.

The charge neutrality condition yields
\begin{equation}
\sum_B q_B (2J_B + 1) k_B^3 \big/ (6\pi^2)
+ \sum_{l=e,\mu} q_l k_l^3 \big/ (3\pi^2)  - \rho_K=0,
\label{neutral}
\end{equation}
where $q_B$ corresponds to the electric charge of baryon species $B$
and $q_l$ corresponds to the electric charge of lepton species $l$. Since
the time scale of a star is effectively infinite compared to the weak
interaction time scale, weak interaction violates strangeness conservation.
The strangeness quantum number is therefore not conserved
in a star and the net strangeness is determined by the condition of
$\beta$-equilibrium which for baryon $B$ is then given by
$\mu_B = b_B\mu_n - q_B\mu_e$, where $\mu_B$ is the chemical potential
of baryon $B$ and $b_B$ its baryon number. Thus the chemical potential of any
baryon can be obtained from the two independent chemical potentials $\mu_n$
and $\mu_e$ of neutron and electron respectively.

\section{Results}

We have always used the parameter set mentioned in the previous
section. The parameter set proposed in \cite{tsushima98} was not
used  since the program fails to converge at high densities for larger bag
radius. As the kaons are only present within these range of densities, we have
opted to use a parametrization with a smaller bag radius.
In what follows we 
look at the neutron star properties that has a phase transition from a hadron
phase described with the QMC model to a phase with hadrons and a kaon
condensate.

\begin{figure}
\begin{tabular}{cc}
\epsfig{file=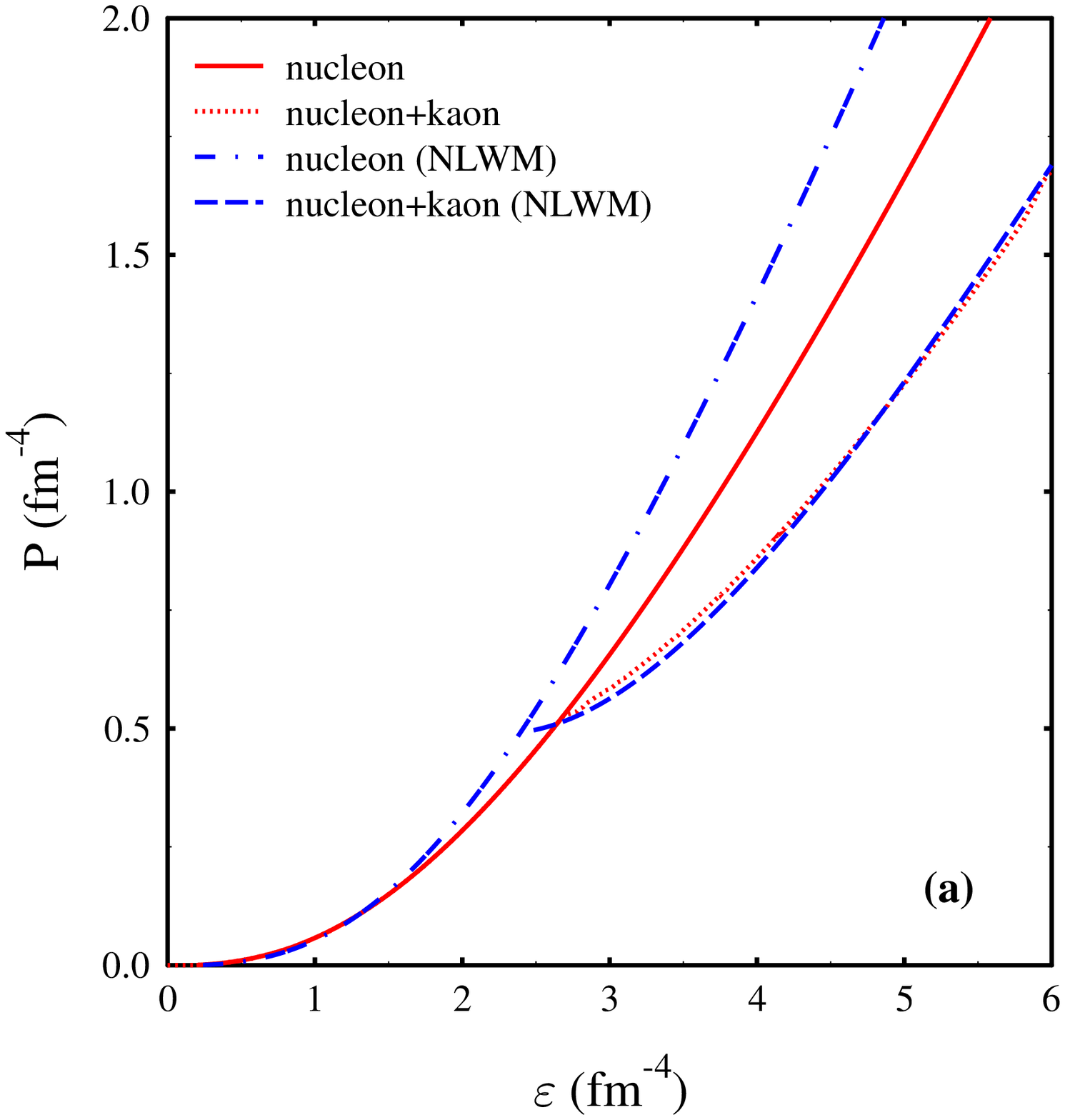,width=7.cm}&
\epsfig{file=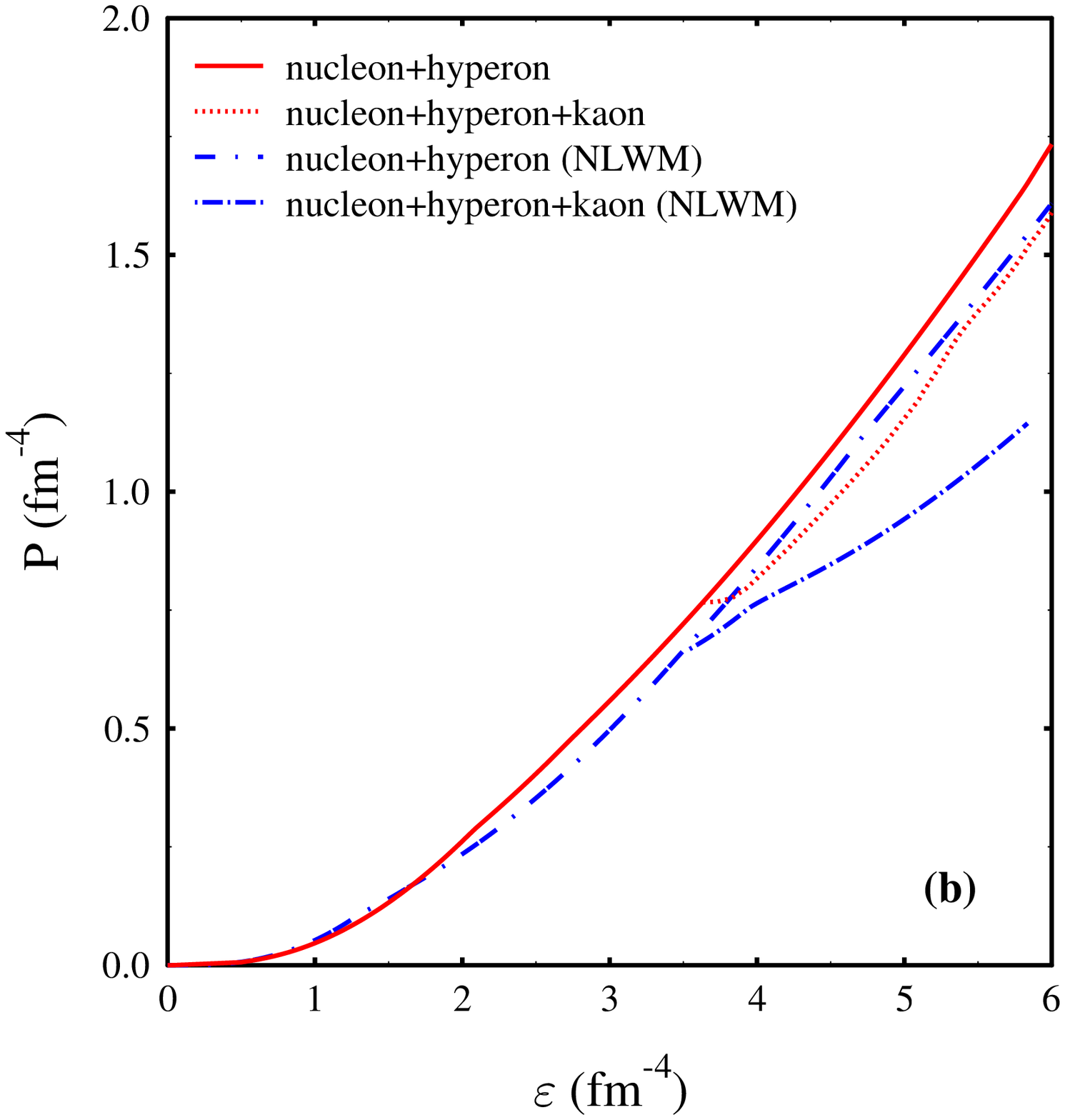,width=7.cm}
\end{tabular}
\caption{Pressure versus energy density for different EOSs}
\label{eos}
\end{figure}

\begin{figure}
\epsfig{file=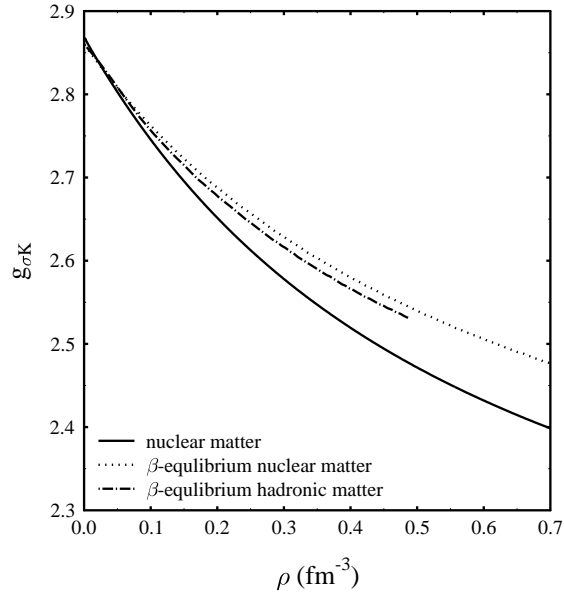,width=8.cm}
\caption{Kaon-$\sigma$ coupling constant, $g_{\sigma K}$ versus density 
obtained within the QMC model}
\label{coup}
\end{figure}

\begin{figure}
\epsfig{file=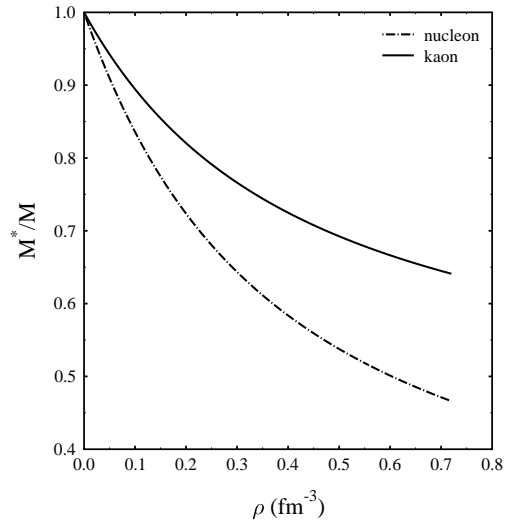,width=7.cm}
\caption{Effective nucleon and kaon masses versus density obtained with the 
QMC model}
\label{effm}
\end{figure}

In figure \ref{eos}a, we show the EOSs obtained  in QMC model
for the nucleons and nucleons plus kaon phase. A kaon condensation is 
possible for densities greater than $3.5\rho_0$, where $\rho_0$ is the 
saturation density. If hyperons are included, fig. \ref{eos}b case, 
kaon condensation appears at even higher densities. In order
to compare the results obtained with the QMC model with the ones
obtained with the NLWM, we also plot the EOS for the NLWM in figures
\ref{eos}a and \ref{eos}b. If no hyperons are present (fig. \ref{eos}a), at 
densities which allow kaon condensation, both models practically coincide, 
although the EOS for the NLWM with nucleons only presents a harder EOS than 
the QMC model. Including hyperons makes the NLWM EOS softer than the QMC one. 
We conclude that the inclusion of strangeness in  a harder EOS, like the
NLWM EOS, has much larger effects. At this point it is worth emphasing that 
in \cite{gl99} all EOS and star properties depend on the choice of the 
potential which fixes the scalar coupling constant in the NLWM, which
is not the case in the present work within the QMC model since this 
coupling is density dependent and an outcome of the calculation depending on 
the bag properties. In figure \ref{coup}
we plot the density dependence of the kaon-sigma coupling for
symmetric nuclear matter, $\beta$-equilibrium nuclear matter and
$\beta$-equilibrium hadronic matter with hyperons. The last curve lies
below the $\beta$-equilibrium nuclear matter, closer to the symmetric
nuclear matter result,  because including hyperons increases the
proton fraction. For the NLWM we have $g_{\sigma K}=1.773$.
In figure \ref{effm} we show the kaon and nucleon effective masses. 
Notice that the mass of the nucleons are more affected by the medium than that
of the kaons, but both decrease quite a lot at the high densities present in
neutron stars.

\begin{figure}
\epsfig{file=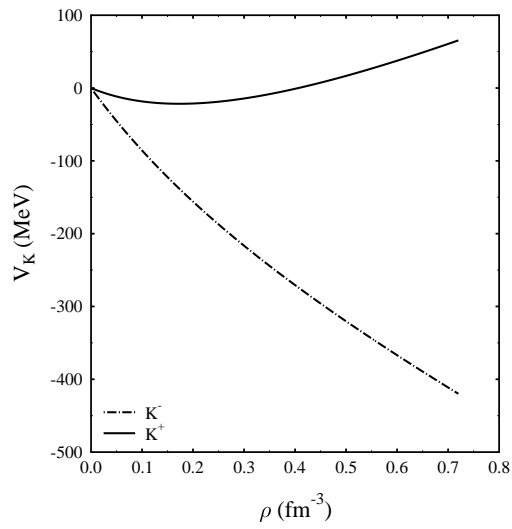,width=7.cm}
\caption{Kaon optical potentials versus density plotted with the QMC model}
\label{disp}
\end{figure}

We define the kaon optical potentials in the present formalism as
$V_{K^\pm}=m^*_K-m_K \pm g_{\omega K} ~\omega_0$. With the present
choice of parameters for the QMC model, the $K^+$ potential is of
the order of -22 MeV and $K^-$ potential amounts to -123 MeV at the
saturation density. In table I we show how these potentials vary
with the bag radius at saturation point. Although the $K^+$
potential does not vary much, the $K^-$ potential changes quite
drastically with the choice of the bag radius.
Another point of interest is the fact that, according
to \cite{gqli}, the $K^+$ feels a weak repulsive potential and the
$K^-$ a strong attractive one. The kaon flow in heavy-ion collisions
was shown to be consistent with the predictions of the chiral
perturbation theory and the values for the $K^+$ and $K^-$
potentials were calculated around 10 and -100 MeV respectively. In
order to obtain a value approximately equal to 10 MeV for the $K^+$ potential, 
we could have rescaled the coupling $g_{\omega K}$ as 1.625 $g_{\omega
K}$ for nucleon bag radius $R_N=0.6$ fm and kaon bag radius $R_K=0.457$ fm.
The authors of \cite{tsushima98}
faced the same problem but they concluded that the rescaling did not
alter any of their qualitative result. Also in \cite{tsushima98} they
pointed that this shortcoming has its origin in the fact that the
bag model does not deal properly with the Goldstone nature of the
$K$-mesons. 
In fig. \ref{disp} we show that, although both potentials are 
attractive at the saturation density, the $K^+$ potential becomes repulsive
at $\sim 2.5\rho_0$. Increasing the kaon bag radius would only give rise to a 
small decrease of the $K^+$ potential, namely for $R_K=0.6$ fm we obtain 
$V_{K^+}=-19$ MeV at saturation density.
For the NLWM calculations we have chosen the $K^-$
potential as -123 MeV and plotted the EOS in fig (\ref{eos}) so that
the comparison with the QMC model would be more appropriate. For the
NLWM we have calculated the $K^+$ potential as being around 16 MeV at 
saturation density.

\begin{table}[h]
\caption{Potential for $K^+$ and $K^-$ at the saturation point for different
bag radius}
\begin{ruledtabular}
\begin{tabular}{ccc}
$R_N$ (fm)& $V_{K^+}$(MeV) & $V_{K^-}$(MeV)\\
\hline
0.6             & -22  & -123  \\
0.8             & -24  & -110  \\
1.0             & -24  & -98
\end{tabular}
\end{ruledtabular}
\end{table}

\begin{figure}
\begin{tabular}{cc}
\epsfig{file=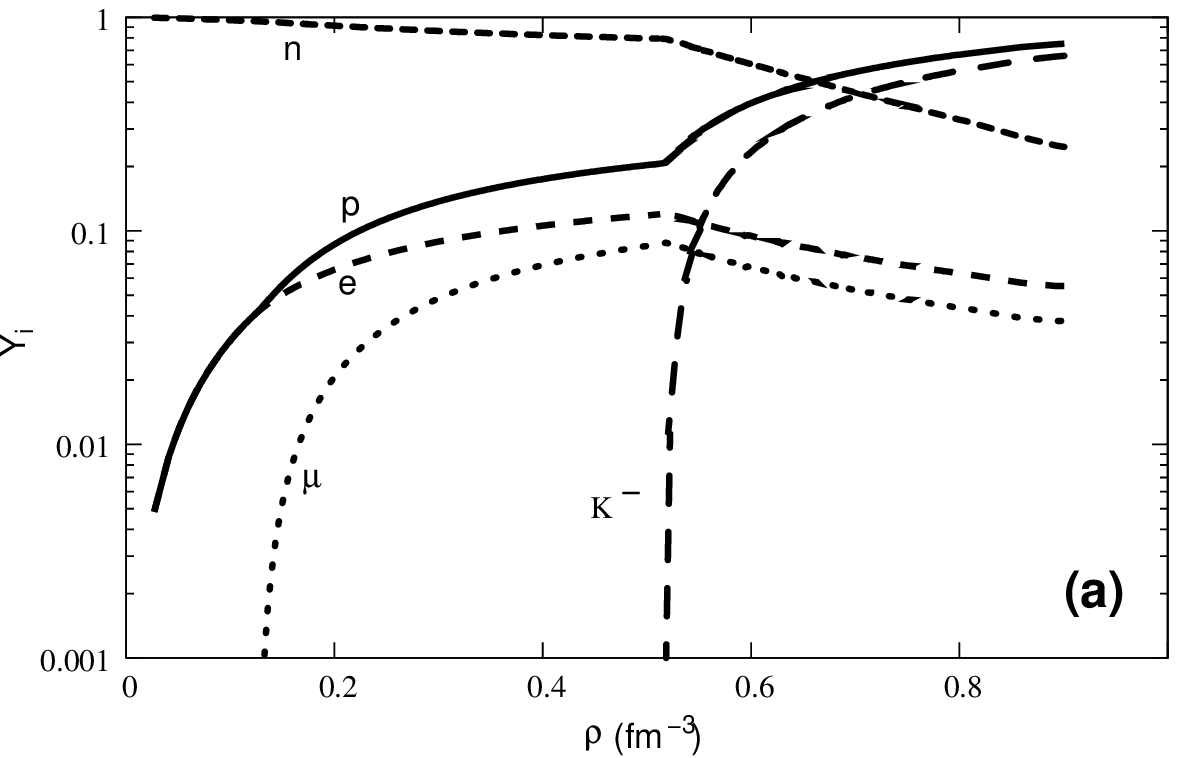,width=7.cm}&
\epsfig{file=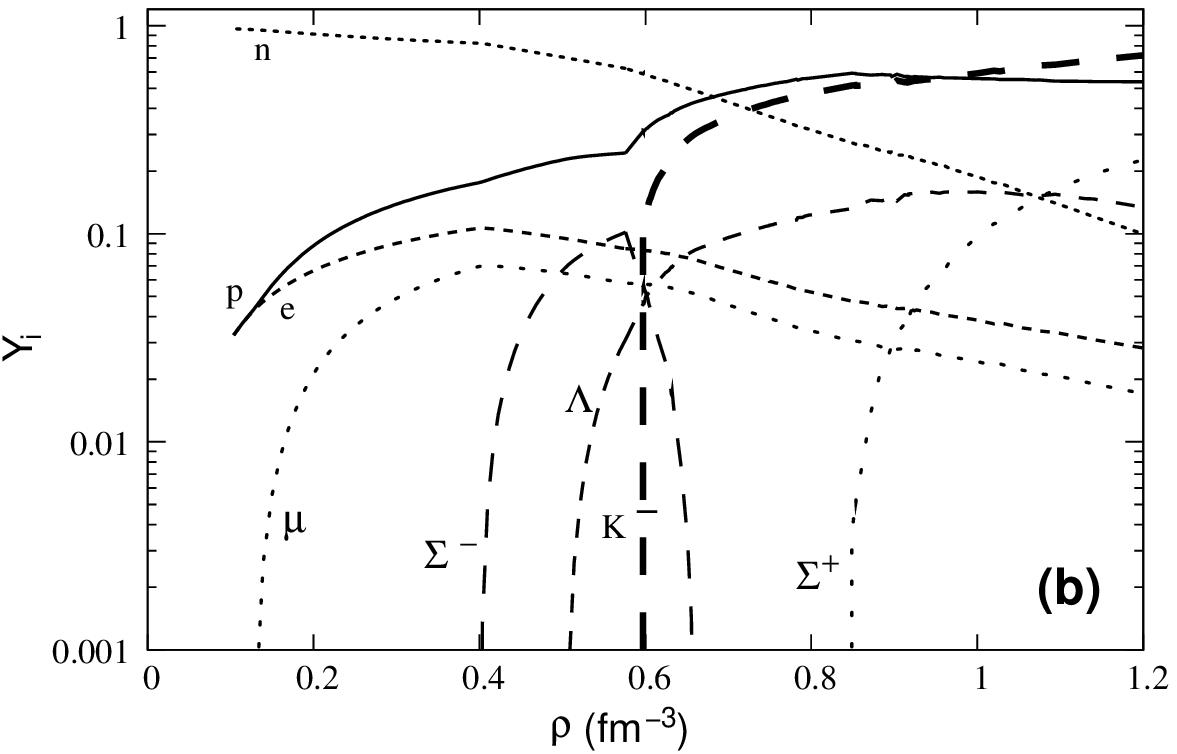,width=7.cm}
\end{tabular}
\caption{Particle population obtained a) without hyperons b) with hyperons}
\label{frac}
\end{figure}

In figure (\ref{frac}a) and (b) we show the particle population for
different densities  including the  possibility of kaon
condensation,  respectively,  without and with hyperons. If we
compare these figures with the figures which display particle
population in \cite{ellis}, where the NLWM was
used for the description of hadrons, we can see that within these models
the electrons and muons disappear at smaller densities than in the present
work. If the hyperons are considered, kaons appear at a higher
density, $\sim 4\rho_0$. Within the present
parametrization the $\Sigma^-$ appears at lower densities than the $\Lambda$
but as soon as the $K^-$ condensates forms, it decreases suddenly.

\begin{figure}
\epsfig{file=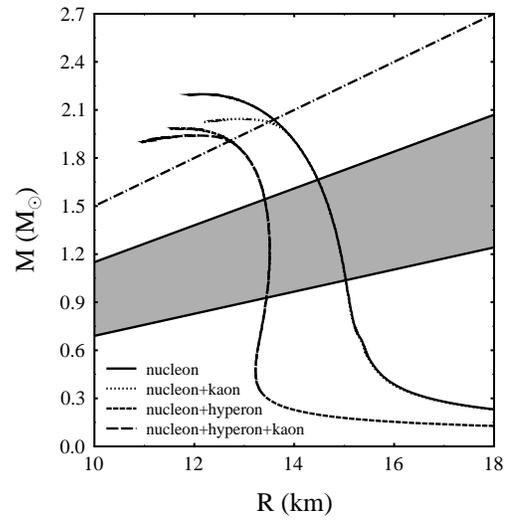,width=8.cm}
\caption{Mass versus radius of the compact stars  for  the QMC model}
\label{tovqmc}
\end{figure}

In figure \ref{tovqmc} we show the mass radius relation of the neutron stars.
In order to describe the crust of the stars we use the
well known BPS EOS calculated in \cite{bps} for very low densities.

Constraints on the mass to radius ration can be obtained from
accurate measurements of the gravitational redshift of spectral
lines produced in neutron star photospheres. A redshift of 0.35 from
three different transitions of the spectra of the X-ray binary
EXO0748-676 was obtained in \cite{cottam}, which corresponds to
$M/R=0.15 M_\odot/km$. Another constraint to the mass to radius
ratio given by $M/R=0.069 M_\odot/km$ to $M/R=0.115 M_\odot/km$ was
determined from the observation of two absorption features in the
source spectrum of the 1E 1207.4-5209 neutron star \cite{sanwal}. In this 
second case, however, the interpretation of the absorption
features as atomic transition lines is controversial. In refs.
\cite{bignami,xu03} the
authors claim that the absorption features are of cyclotron nature, which
would make the related constraints unrealistic. In
figure \ref{tovqmc} we have added the lines corresponding to that
constraints. We can see that all the curves presented in this work
are consistent with the constraints.

In table II we show the stellar properties for the EOS discussed
above, namely, the maximum gravitational and baryonic masses, the
central energy density and the radius of the star with the maximum
mass.
\begin{table}[ht]
\caption{Hadronic and hybrid star properties for the EOSs
described in the text}
\begin{ruledtabular}
\begin{tabular}{lccccccc}
type & hadron model & $M_{max}(M_\odot)$ &
$M_{b~max}(M_\odot)$ & $R$ (Km) & $\varepsilon_0$ (fm$^{-4}$)\\
\hline
np                     & QMC  & 2.20 & 2.59 & 12.15 & 5.74 \\
np+kaon                & QMC  & 2.05 & 2.36 & 12.94 & 4.80 \\
np+hyperons            & QMC  & 1.98 & 2.28 & 11.76 & 5.87 \\
np+hyperons+kaon       & QMC  & 1.94 & 2.27 & 12.00 & 5.47 \\
np                     & NLWM & 2.40 & 2.90 & 12.23 & 5.44 \\
np+kaon                & NLWM & 2.08 & 2.42 & 13.21 & 4.32 \\
np+hyperon             & NLWM & 1.90 & 2.09 & 11.94 & 5.49 \\
np+hyperon+kaon        & NLWM & 1.84 & 2.04 & 12.52 & 5.03
\end{tabular}
\end{ruledtabular}
\end{table}
From table II and figure \ref{tovqmc} one can observe that
the maximum gravitational mass always decreases when kaons are
introduced, a consequence of the fact that the EOS for 
$\beta$-equilibrium matter is softer with the presence of the
strangeness. In our calculations the maximum gravitational mass in the QMC 
model always decreases with the inclusion of kaon by a small amount, $\sim 
0.15M_\odot (\sim 0.05M_\odot)$ without (with) hyperons. Different 
NLWM models were mentioned in \cite{gqli}, where the decrease obtained with the
introduction of kaon condensate was of the order of 0.4 $M_{\odot}$,
a number that can change by about 20\% when different nuclear EOS
are used. This is in agreement with the results we have obtained
within a NLWM parametrization for which the decrease is about 0.32
$M_{\odot}$. The difference between NLWM and QMC  model is due to the
fact that the QMC nucleon EOS is softer and, therefore the effect of
including kaons is not so large as in the NLWM.
The maximum baryonic mass follows the same trend as the gravitational mass.
The radius always increases with the introduction of the
kaon condensate and the central energy density decreases, again due to a   
softer EOS. Although a nuclear compact star within the NLWM has a higher mass 
than within the QMC model, once we consider the possibility of kaon 
condensation both models give similar results. Including hyperons together 
with a kaon condensation decreases slightly the maximum gravitational mass, 
as expected because we are using a softer EOS, but it has also the effect of 
decreasing the radius  as compared with nucleons plus kaons
and increasing the central energy density.
If we compare a star buit from a EOS with nucleons plus hyperons with a 
star containing nucleons plus kaons, we observe that they both present
similar maximum gravitational masses, slightly larger for the second case, 
the radius and central energy density of the corresponding stars are quite 
different: in a nuclear star with kaon condensation the maximum central energy 
density is quite low and therefore the radius large.
This is true both within the QMC and the NLWM models.

In summary, we have investigated  within the QMC model the consequences of 
including kaon condensation  on the EOS and related compact 
star properties. Within the QMC the kaon-sigma coupling constant is
obtained self-consistently with the calculation of the kaon
effective mass. This definition of the kaon-meson couplings gives rise to a 
$K^+$ optical potential which is slightly attractive for densities
smaller than $2.5\rho_0$, contrary to what was predicted by a chiral model 
\cite{gqli}. Some conclusions with respect to compact star properties described
within the present formalism are:
a) kaon condensation  makes the EOS softer and reduces the  maximum allowed 
stellar masses . A similar effect is obtained with the inclusion
of hyperons. Including only kaons does not reduce the maximum mass so much 
as including only hyperons, however, the radius of the maximum mass 
configuration is larger and its central energy density smaller;
b) these same conclusions were confirmed within the NLWM although  the
mass reduction due to the inclusion of strangeness is larger in
this model;
c) the inclusion of both hyperons and kaon condensation makes the EOS even 
softer, decreasing the maximum star mass allowed, but also increasing the 
central energy density and decreasing the radius with respect to the stars 
with kaon condensation only;
d) finally the inclusion of strangeness in the EOS of compact stars within the 
QMC model does not have such a strong effect as when the NLWM models are used.
The existing constraints on the mass to radius 
ratio do not exclude any of the EOSs discussed in the present work.

\section*{ACKNOWLEDGMENTS}

This work was partially supported by Capes (Brazil) under process BEX
1681/04-4, CAPES (Brazil)/GRICES (Portugal) under project 100/03 and
FEDER/FCT (Portugal) under the project  POCTI/FP/FNU/50326/2003.
D.P.M. would like to thank the friendly atmosphere at the Reserch Centre
for Theoretical Astrophysics in the Sydney University, where this
work was partially done.

\newpage

\end{document}